\documentclass[aps,amssymb,amsmath,prl,twocolumn,showpacs,superscriptaddress]{revtex4-1}
\usepackage{graphicx}
\usepackage{bm}
\usepackage{xcolor}
\usepackage{graphicx}
\newcommand{\CP}[2]{\Pr\Big( #1 \Big\vert #2 \Big)}

\begin{document}
\title{Stochastic quorum percolation and noise focusing in neuronal networks}

\author{Javier G. Orlandi}
\email{javier.orlandigomez@riken.jp}
\affiliation{Department de F\'isica de la Mat\`eria condensada, Universitat de Barcelona, E-08028 Barcelona, Spain}
\affiliation{RIKEN Center for Brain Science, Wako-shi, 351-0198 Saitama, Japan}
\author{Jaume Casademunt}
\affiliation{Department de F\'isica de la Mat\`eria condensada, Universitat de Barcelona, E-08028 Barcelona, Spain}
\affiliation{Universitat de Barcelona Institute of Complex Systems (UBICS), Universitat de Barcelona,  E-08028 Barcelona, Spain}

\pacs{87.19.lj,64.60.ah,87.18.Sn}

\begin{abstract}
Recent experiments have shown that the spontaneous activity of young dissociated neuronal cultures can be 
described as a process of highly inhomogeneous nucleation and front propagation due to the localization of noise activity, i.e., noise focusing. However, the basic understanding of the mechanisms of noise build-up leading to the nucleation remain an open fundamental problem. Here we present a minimal dynamical model called stochastic quorum percolation that can account for the observed phenomena, while providing a robust theoretical framework. The model reproduces the first and second order phase--transitions of bursting dynamics and neuronal avalanches respectively, and captures the profound effect metric correlations in the network topology can have on the dynamics. The application of our results to other systems such as in the propagation of infectious diseases and of rumors is discussed.
\end{abstract}

\date{\today}

\maketitle

The spontaneous activity of young dissociated neuronal cultures is characterized by bursts of high-frequency collective activity followed by silent periods, with little activity~\cite{Maeda.1995,Eckmann.2007}. It has recently been shown that burst emergence can be explained by a process of nucleation and front propagation~\cite{Orlandi.2013lxs}. Multiple nucleation sites coexist in a given culture, i.e., zones of high nucleation probability, where a burst can develop and propagate. The presence of nucleation sites and its spatial heterogeneity 
arise due to noise focusing~\cite{Orlandi.2013lxs}, a symmetry-breaking mechanism that amplifies the quenched disorder in the network wiring, modifying the transport properties of the spontaneous activity. The resulting coarse-grained physical picture is that noise flows anisotropically through the metric space where the network is embedded as it is amplified by the integrate-and-fire dynamics of its nodes, concentrating at some specific locations. 

While this scenario has been recently understood at a mesoscopic level~\cite{Orlandi.2017586}, the mechanisms of dynamical and topological amplification~\cite{Orlandi.2013lxs} that are responsible for this phenomenon are difficult to grasp at a quantitative level. To gain insight into this central point and in particular to elucidate the role of metric connectivity correlations (which play a major role in the behavior of neuronal cultures~\cite{Hernandez-Navarro.2017}) we present here a minimal statistical model, called stochastic quorum percolation (SQP), that captures the generic features of the noise focusing mechanism by extending the previous theory of quorum percolation~\cite{Cohen.2010d0q} into a fully dynamical model, while simultaneously providing a natural connection with other theoretical frameworks such as directed, compact percolation, branching processes and cellular automata~\cite{Lubeck.2004}.

A few years ago, the concept of quorum percolation (QP) was introduced~\cite{Cohen.2010d0q} to describe the collective dynamics of neuronal networks under external stimulation~\cite{Soriano.200883l}. In these experiments, the network connectivity is weakened with different drugs and an external current is applied to the system to study its response. QP is an extension of percolation theory in which a node requires a minimum quorum of $m$ simultaneous inputs to become active. For a given initial fraction $f$ of active nodes, the quorum condition is checked iteratively until no more nodes can be activated. QP is characterized by a discontinuous phase transition, where a small increase in $f$, generates a large response in the size of the final active fraction $\Phi$. QP is similar to bootstrap~\cite{Tlusty.2009} and k-core percolation~\cite{Dorogovtsev.2006}, although several differences exist, specially in the presence of disorder~\cite{Renault.2014,Monceau.2016}.

In the absence of any external stimuli, however, neuronal systems are spontaneously active, and the propagation of neuronal activity is often described in terms of avalanches~\cite{Beggs.2003c9f}, where essentially any neuron firing can trigger a firing on its neighbours with probability $p_1$. By extension, a neuron receiving $k$ inputs will fire with probability $p_k=1-(1-p_1)^k$. On the other hand, in QP a neuron fires if, and only if, it receives $m$ inputs, $p_k=\Theta(m-k)$, where $m$ is called the quorum threshold. However, real neurons are driven by noise, and always have a finite spontaneous firing rate. In the stochastic quorum percolation (SQP) model we assume that the internal noise of the neuron is characterized by a Poisson process with a given rate $\lambda$ (a shot noise). This noise model is inspired on minis (spontaneous miniature post synaptic currents), which in sparse cultures have the same strength as evoked currents~\cite{Cohen.2009zlp,Cohen.2011ejc}. This description is also equivalent with considering an unobserved population projecting onto the observed network and firing with Poisson statistics. Hence for simplicity, we assume that each discharge of our shot noise has the same effect as receiving one input from another neuron.
Within this description, the probability that a neuron fires spontaneously within a time window $\Delta t$ is
\begin{eqnarray}
  p_0 = e^{-\lambda\Delta t}\sum\limits_{i=m}^\infty\frac{(\lambda\Delta t)^i}{i!}={\cal P}(m,\lambda\Delta t),
\end{eqnarray}
where ${\cal P}$ is the regularized gamma function, and we choose $\Delta t$ as the characteristic integration time of synaptic currents $\Delta t\approx 20ms$. Hence, the spontaneous firing frequency of a neuron is $\omega_0\approx p_0/\Delta t$. 
Accordingly,
an isolated neuron spontaneously fires when it accumulates the required quorum of $m$ shots from a Poisson process with rate $\lambda$ within $\Delta t$. The case where a neuron is receiving $k$ external inputs within $\Delta t$ is equivalent to lowering the firing quorum from $m$ to $m-k$, i.e.,
\begin{eqnarray}
  p_k = {\cal P}(m-k, x),
\end{eqnarray}
where we define $x\equiv\lambda\Delta t$ as the noise strength. In the limit of large $m$ and \emph{weak} noise $x\to0$, we recover the expression of the original QP model $p_k=\Theta(m-k)$. 

At its essence, SQP is a dynamical process. It has spontaneous activations that can induce other activations throughout the network, and given enough inputs (quorum), provides a deterministic response. This minimal set of features can be related, one by one, to classical models of non-equilibrium phase transitions: its dynamical nature is that of directed percolation~\cite{Hinrichsen.2000}; the spontaneous activations are related to the presence of external fields in non-equilibrium systems, e.g., the spontaneous generation of new particles in the pair-contact process~\cite{beck.2002}; and the quorum condition is akin to the presence of an upper absorbing state in compact directed percolation and in the Domany-Kinzel model~\cite{Domany.1984,Essam.1999}.

The temporal evolution of SQP can be described as a Markov process within a network. The network is characterized by its size $N$ and its adjacency matrix A, whose entries $A_{ij}=1$ denote the existence of a connection from node $i$ to node $j$. And the state of each node in the system $S_i$ is either active $S_i=1$ or inactive $S_i=0$. The evolution of the system is characterized by its transition probabilities 
\begin{eqnarray}
  \CP{S_i(t+1)=1}{\vec S(t)}={\cal P}(m-k_i(t),x),
  \label{eq:markov}
\end{eqnarray}
where $\vec S(t)=(S_1(t),\ldots,S_N(t))$ and $k_i(t)=\left(A^\mathsf{T}\vec S(t)\right)_i=\sum_{j}A_{ji}S_j(t)$, i.e., the number of active inputs of $i$ at time $t$.

\begin{figure}
  \centering
  \includegraphics{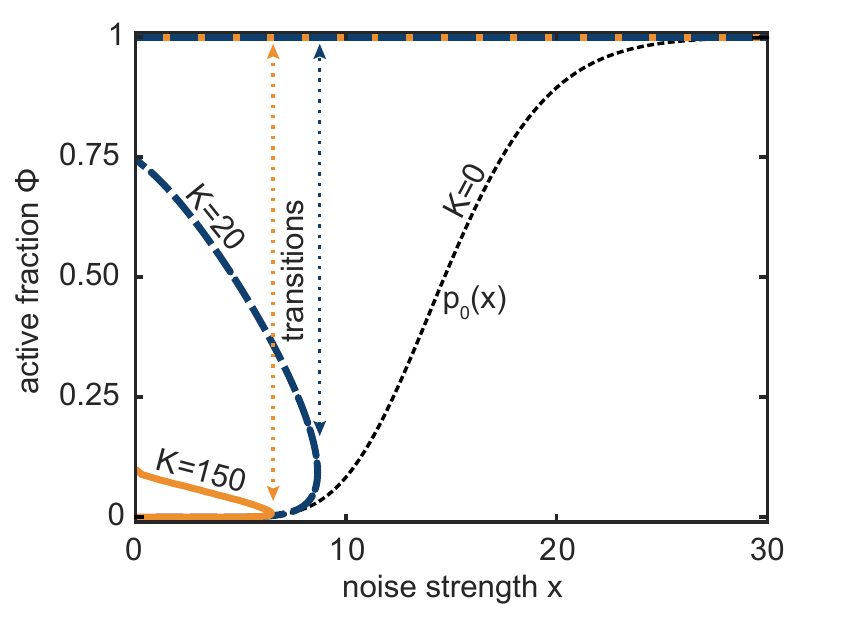}
  \caption{Bursting transition.
    In orange and blue (solid, dashed line), the stable solutions of eq.~\eqref{eq:steadystatefull} for $m=15$ and $K=150$ and $K=20$ for different levels of noise $x$. Dotted line, solution for a fully disconnected network, i.e., $K=0$. For low levels of noise (even at $x=0$), $\Phi$ has three possible solutions, the middle one being unstable. There exists a critical value of $x$, $x^*$ (as long as $m<K$) where the two lower branches merge that characterize the discontinuous bursting transition.
    \label{fig:subquorum:transition}
  }
\end{figure}

The model can be analyzed at the mean--field level in the limit of a maximally entropic infinite random graph with fixed in and out degree distributions, $p(K_i)=p(K_o)=p(K)$, i.e., $N\to\infty$, $\langle CC\rangle\to0$ (average clustering coefficient).
We characterize the system by the fraction of active nodes at time $t$, $\Phi(t)=N^{-1}\sum_iS_i(t)$ and look for the steady state solution
$\Phi(t+1)=\Phi(t)$. 
Note from eq.~\eqref{eq:markov} that taking into account all possible combinations of active inputs we can write
\begin{eqnarray}
  \Phi = \sum_Kp(K)\sum\limits_{n=0}^K{K\choose n}{\cal P}(m-n,x)\Phi^n(1-\Phi)^{K-n},
  \label{eq:steadystatefull}
\end{eqnarray}
which can 
be solved numerically~\footnote{${\cal P}(j,x)$ can be analytically continued to any real value of $j$, and in particular it vanishes for negative integers.}. 
Note that ${\cal P}$ has a sigmoid shape, and for all relevant cases, eq.~\eqref{eq:steadystatefull} has either one or three solutions (see Fig.~\ref{fig:subquorum:transition}). Without loss of generalization, we will assume a connectivity distribution $p(K)=\delta(K-\bar K)$ from now on.

For $m<\bar K$ (which is met for any real system, i.e., each node has more connections than the quorum needed to activate), and low levels of noise $x$, eq.~\eqref{eq:steadystatefull} has three solutions, two stable and one unstable, even without noise ($x=0$). The system can be found in two regimes: one with very low activity, where $\Phi\sim p_0$, and the system is essentially inactive, and one where all the nodes are active, $\Phi=1$, signaling bursting behavior. There exists a critical noise value $x^*$, for which the low activity solution and the unstable solution merge, hence defining  
a discontinuous phase--transition. For any $x>x^*$, the whole system is active. This transition to $\Phi=1$ is equivalent to the emergence of a burst in real neuronal networks, as we will see later on. $\Phi=1$ is an absorbing state, and the inclusion of a mechanism similar to short-term synaptic depression is needed to get the inverse transition (from bursting to non-bursting). Note that if the system size is finite, for $x<x^*$ there exists a finite probability per unit time of transition between the states, so the system will eventually reach the absorbing state, i.e., it will burst in a finite time.

In the low activity regime, the steady state of the system is characterized by the presence of cascades of activity, or avalanches, defined as a temporal sequence of causally connected activations. Similarly to what is done in branching processes and studies of criticality in similar systems~\cite{Beggs.2003c9f,Kinouchi.2006,Costa.2015d2m}, we can define a local branching ratio as
\begin{eqnarray}
  \sigma = \bar K\sum\limits_{n=0}^{\bar K-1}{\bar K-1\choose n}\phi^n(1-\phi)^{\bar K-n-1}{\cal P}(m-n-1, x),
  \label{eq:branchingratio}
\end{eqnarray}
for a network with fixed connectivity $\bar K$, where
$\sigma$ denotes the average number of nodes that a given active node will activate in a posterior time step. If we impose $p_0=0$, which corresponds to having separation of time-scales between random activations and activity propagation, $\sigma=1$ marks the presence of sustained activity, i.e., the active phase. 
There exists a critical level of noise, $x^\dagger$, where $\sigma=1$ and $x^\dagger$ marks a continuous, second-order phase-transition. 
For $x<x^\dagger$, there is an absorbing state with $\Phi=0$, the inactive phase, and any cascade of activity will eventually die out. 
From eq.~\eqref{eq:branchingratio} we obtain in the inactive phase that $\sigma=\bar K{\cal P}(m-1, x)$, hence $
x^\dagger$ is easily obtained from the relation $1/\bar K={\cal P}(m-1, 
x^\dagger)$, i.e., $p_1=1/\bar K$, as in most kinds of percolation on a Bethe lattice~\cite{Albert.2002}. In fact, the model belongs to the same universality class as directed percolation.

\begin{figure}
  \centering
  \includegraphics{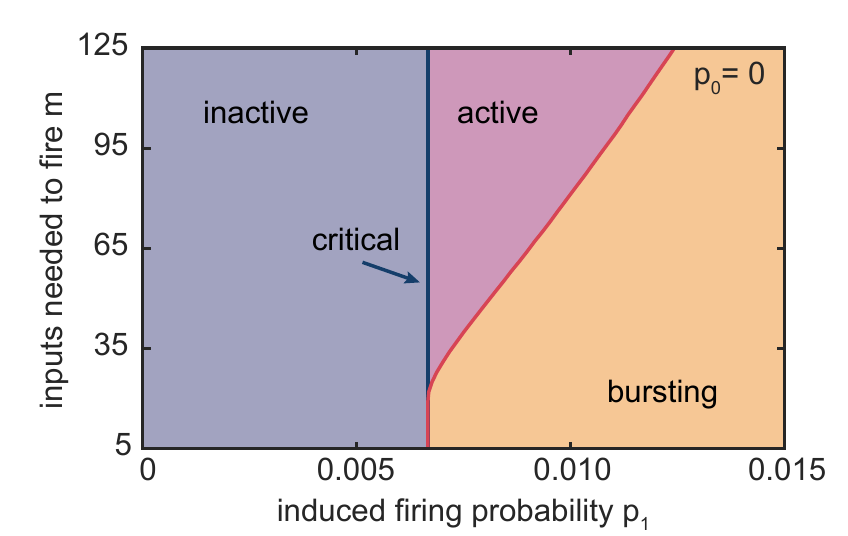}
  \caption{Phase Diagram for a network with fixed connectivity $\bar K=150$ within mean--field. In the absence of spontaneous activity $p_0=0$. The vertical (blue) line at $p_1=1/\bar K$ marks a continuous phase--transition between the inactive and active phases. The other (red) line marks a discontinuous phase--transition into the bursting phase, where all the nodes are active. Note that for low $m$ both transitions occur at the same $p_1$.
    \label{fig:subquorum:phaseDiagram}
  }
\end{figure}

The presence of spontaneous activity, $p_0\neq0$, however, destroys the lower absorbing state ($x>0$) and consequently the transition~\cite{beck.2002}. Regardless, one can still define a critical noise value $x^\dagger$ with maximum susceptibility $\chi=d\Phi/dp_0$ that defines a nonequilibrium Widom line~\cite{Williams-Garcia.2014,vg}. The mean--field phase--diagram for a given network connectivity is shown in Fig.~\ref{fig:subquorum:phaseDiagram}, for the particular case of $p_0=0$.  

Although the previous mean--field approximation is only valid for infinite random graphs, a similar phase--diagram can be constructed for real networks. However, several differences exist. In a finite network, even for the case of a random graph, given that the system is driven by noise, it can reach the upper absorbing state ($\Phi=1$) even for low levels of noise. It only needs a fluctuation in the number of active nodes $\Phi$ big enough to reach the unstable solution of eq.~\ref{eq:steadystatefull} (see Fig.~\ref{fig:subquorum:transition}). In this situation the lifetime of the fluctuations of the system, as well as the temporal correlation in the activity of the nodes, is going to depend strongly on the precise network topology, as we will now see.

To show the effects network of topology on the dynamics we will restrict ourselves to the study of the metric networks presented in Ref.~\cite{Orlandi.2013lxs} that reproduce the structure of dissociated neuronal cultures. These networks are obtained as follows: neuronal bodies (nodes) are placed at random in a substrate, usually a square of size $L$ with periodic boundary conditions until a desired density is reached ($\rho=100-1000~\mathrm{neurons/mm^2}$). Then, an axon is grown from each body as a biased random--walk that results in almost straight axons with a characteristic length $\ell_a\sim1~\textrm{mm}$ and finally, the dendritic tree of a neuron is modeled as a circular area around its body with a characteristic radius of $r_d\sim0.15~\textrm{mm}$. A connection between two neurons $i\to j$ is created with probability $\alpha$ whenever the axon of neuron $i$ intersects with the dendritic tree of neuron $j$.

This procedure generates networks with in and out --degree distributions characterized by the distributions of dendritic and axonal sizes respectively (we usually choose a Gaussian one for the first and a Rayleigh for the latter), and with a clustering coefficient $CC$ that depends on $\alpha$ and $\rho$. Typically $\langle CC\rangle\sim0.25$. These networks present connectivity correlations that decay exponentially with a characteristic length of $\ell_c=0.26~\textrm{mm}$, and can be seen as 
random graphs at small distances $r<0.15~\textrm{mm}$ but are highly directional at long distances due to the particular morphology of neurons, with all their output connections found in a narrow area around their axons. To study the impact that the underlying metric correlations have on the network dynamics, for each constructed network we generate a maximally entropic conjugate network with the constraint of keeping the same in and out -degree distributions via a random swap of links between unconnected neuronal pairs.

\begin{figure}
  \centering
  \includegraphics{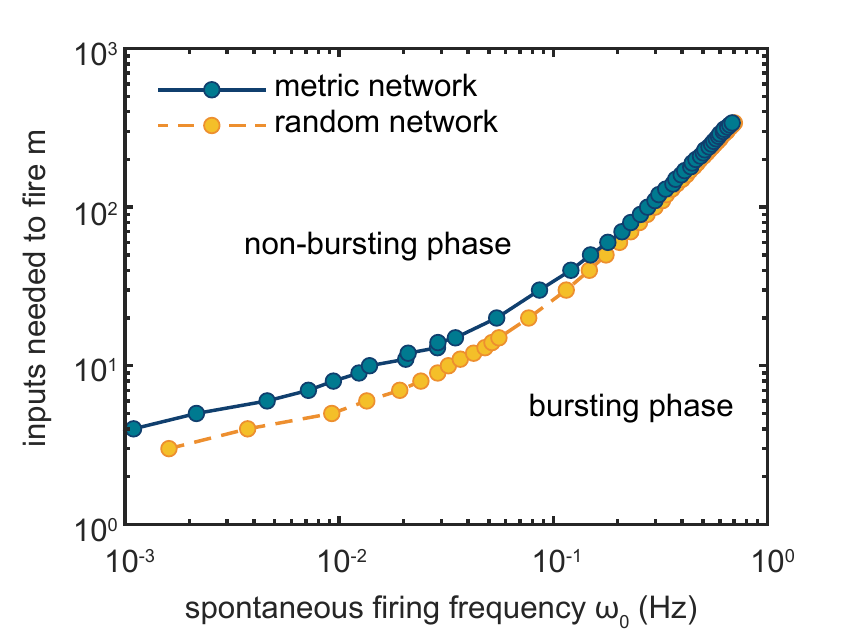}
  \caption{Phase Diagram for a simulated neuronal network and its randomized conjugate (see text) in log--log scale. $\rho=500~\mathrm{neurons/mm^2}$, $N=8000$, $\bar K=185$, $\langle CC\rangle=0.26$. The corresponding lines mark the approximate transition boundary into the bursting phase for each network configuration. As shown, the metric network topology always favors the transition, occurring at a lower spontaneous frequency for any quorum threshold $m$.
  \label{fig:subquorum:phaseDiagramSimulated}
  }
\end{figure}

As we have discussed previously, in finite systems with spontaneous activity (which is the case for any real system), the continuous phase--transition does not really exist and the discontinuous transition is not marked by a sharp line and instead is associated to a characteristic first--passage time, that we will call the bursting time, i.e., the average time it takes the system to reach the bursting phase, i.e., the upper absorbing state. In Fig.~\ref{fig:subquorum:phaseDiagramSimulated} we characterize the transition by the smallest spontaneous firing rate $\omega_0=p_0/\Delta t$ required for the system to reach the bursting phase within a fixed, arbitrarily long time (5000 s). 
This figure clearly shows the important effect of network topology in the dynamics, where the presence of clustering and higher--order correlations in the network structure, favors the presence of the bursting phase, shifting the transition line to smaller spontaneous firing rates.

\begin{figure}
  \centering
  \includegraphics{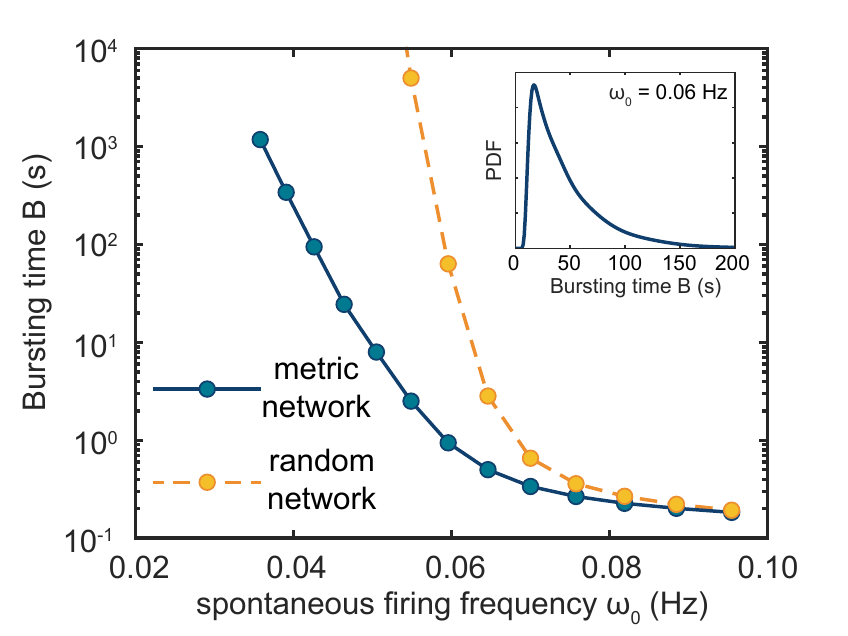}
  \caption{Characteristic bursting times. Same network as in Fig.~\ref{fig:subquorum:phaseDiagramSimulated}, with $m=15$. For high levels of noise the network structure becomes irrelevant and both networks show similar bursting time. For low values of the noise, however, in the absence of network correlations the bursting time increases much faster. Inset: probability distribution function of bursting times in the metric case at $\omega_0=0.06~\mathrm{Hz}$.
  \label{fig:subquorum:ignitionTimes}
  }
\end{figure}

This effect is more dramatically observed in the dependence of the bursting time with the spontaneous firing rate for a given quorum threshold $m$ (see Fig.~\ref{fig:subquorum:ignitionTimes}). For high spontaneous firing rates, the bursting time  is insensitive to the topology. For low rates, however, it has a strong dependence, increasing at a much faster rate in the random case than in the metric one. The range of frequencies where the system can reach the bursting phase in a realistic time is largely expanded in the metric network.

The strong dependence on the bursting time with the metric correlations 
of the network is an essential feature of the noise focusing mechanism, as pointed out in Ref.~\cite{Orlandi.2013lxs}. The non--linear behavior of the induced firing probability function ${\cal P}$, results in lower requirements of simultaneous inputs to reach the bursting phase in regions of high input connectivity and clustering, for which the noise is effectively stronger. In a network embedded in a metric space, once the bursting phase is reached within a finite region the active phase will spread to the whole system (as long as $\bar K/2>m$~\cite{Tlusty.2009}), regardless of its size.

The SQP model provides an appropriate description of neuronal dynamics in terms of percolation concepts, providing a simplified and unified framework to account for both bursting dynamics and neuronal avalanches (power--law statistics of cascades of induced firings). The two types of dynamics coexist within the same model, and bursting behavior can appear with either critical or supercritical avalanches.

For simplicity, and with the aim of identifying general mechanisms that go beyond the case of neuronal systems, we have not considered the effects of synaptic depression, facilitation, inhibition or variable synaptic weights, but these are straightforward additions to the model, since all of them can be associated to changes in ${\cal P}$. 
Note that with the addition of synaptic depression or inhibition, the model can easily exhibit self--organized criticality, as in Ref.~\cite{Levina.2009}.

Neuronal avalanches are characteristic of many neuronal systems, and have been described in a myriad of systems within the theory of branching processes~\cite{Beggs.2003c9f,Pasquale.20085f,Tetzlaff.2010y3q,Priesemann.2014,Yaghoubi.2018}, and recently with directed percolation\cite{Carvalho.2020}. But even more ubiquitous is the presence of bursts, which are a direct consequence of the integrate-and-fire dynamics. For sufficient activity the system will exhibit a discontinuous transition, not just super-critical avalanches, as predicted by simpler models. Critical neuronal avalanches are believed to be a desirable feature~\cite{Beggs.20083u}, and there have been many attempts to describe mechanisms by which neuronal systems self-organize towards that behavior~\cite{Levina.2009,Costa.2015d2m,Papa.2017fe}. System-wide bursts, on the other hand, are often linked to epileptic behavior and run-away excitation~\cite{Touboul.2011}, and neuronal systems have to implement mechanisms to prevent, stop or reduce them. In particular, networks with balanced excitatory and inhibitory connectivity, often show smaller and heterogeneous bursting activity, both in \textit{in vitro} and \textit{in vivo} preparations.

The behavior presented here is not exclusive of neuronal systems, as any process running on a network of integrate-and-fire elements with non--linear summation of probabilities should operate in a similar way. Our model could easily be adapted to disease--spreading processes for diseases where the probability of contagion can quickly rise or without separation of time-scales\cite{vg}; or to rumor spreading with the illusion of truth effect~\cite{Moons.2009}, where the bombardment of false information in a short time window can lead to false beliefs and further spreading.

\begin{acknowledgments}
We thank J. Davidsen and R. Williams-Garcia for useful discussions. We thank financial support of MINECO under projects FIS2013-41144-P, FIS2016-78507-C2-2-P, and from Generalitat de Catalunya under project 2014-SGR-878.
\end{acknowledgments}

\bibliography{sqp}





\end{document}